\documentclass[12pt,twocolumn]{aastex63m}
\usepackage{graphicx}
\usepackage{longtable}
\def\agile{{\it AGILE} }
\def\agilep{{\it AGILE}}
\def \chf{CHIME/FRB }
\def \sgr{SGR 1806-20 }
\def \sgrp{SGR 1806-20}

\def \be {\begin{equation}}
\def \en {\end{equation}}

\def \fc {}
\def \fv {}
\def \fvv {}
\def \fvvv {}
\def \mt {}
\def \mtt {}

\begin{document}
\title{\large{\agile Observations of Two Repeating Fast Radio Bursts with Low
Intrinsic Dispersion Measures}}
\author{C. Casentini}
\email{claudio.casentini@inaf.it}
\affiliation{\scriptsize INAF/IAPS, via del Fosso del Cavaliere 100, I-00133 Roma (RM), Italy}
\affiliation{\scriptsize INFN Sezione di Roma 2, via della Ricerca Scientifica 1, I-00133 Roma (RM), Italy}

\author{F. Verrecchia}
\email{francesco.verrecchia@inaf.it}
\affiliation{\scriptsize SSDC/ASI, via del Politecnico snc, I-00133 Roma (RM), Italy}
\affiliation{\scriptsize INAF -- Osservatorio Astronomico di Roma, via Frascati 33, I-00078 Monte Porzio Catone (RM), Italy}

\author{M. Tavani}
\email{marco.tavani@inaf.it}
\affiliation{\scriptsize INAF/IAPS, via del Fosso del Cavaliere 100, I-00133 Roma (RM), Italy}
\affiliation{\scriptsize Universit\`a degli Studi di Roma Tor Vergata, via della Ricerca Scientifica 1, I-00133 Roma (RM), Italy}

\author{A. Ursi}
\affiliation{\scriptsize INAF/IAPS, via del Fosso del Cavaliere 100, I-00133 Roma (RM), Italy}
\author{L.A. Antonelli}
\affiliation{\scriptsize INAF -- Osservatorio Astronomico di Roma, via Frascati 33, I-00078 Monte Porzio Catone (RM), Italy}
\author{A. Argan}
\affiliation{\scriptsize INAF/IAPS, via del Fosso del Cavaliere 100, I-00133 Roma (RM), Italy}
\author{G. Barbiellini}
\affiliation{\scriptsize Dipartimento di Fisica, Universit\`a di Trieste and INFN, via Valerio 2, I-34127 Trieste (TR), Italy}
\author{A. Bulgarelli}
\affiliation{\scriptsize INAF/OAS, via Gobetti 101, I-40129 Bologna (BO), Italy}
\author{P. Caraveo}
\affiliation{\scriptsize INAF/IASF, via E. Bassini 15, I-20133 Milano (MI), Italy}
\affiliation{\scriptsize INFN Sezione di Pavia, via U. Bassi 6, I-27100 Pavia (PV), Italy}
\author{M. Cardillo}
\affiliation{\scriptsize INAF/IAPS, via del Fosso del Cavaliere 100, I-00133 Roma (RM), Italy}
\author{P.W. Cattaneo}
\affiliation{\scriptsize INFN Sezione di Pavia, via U. Bassi 6, I-27100 Pavia (PV), Italy}
\author{A. Chen}
\affiliation{\scriptsize School of Physics, Wits University, Johannesburg, South Africa}
\author{E. Costa}
\affiliation{\scriptsize INAF/IAPS, via del Fosso del Cavaliere 100, I-00133 Roma (RM), Italy}
\author{I. Donnarumma}
\affiliation{\scriptsize ASI, via del Politecnico snc, I-00133 Roma (RM), Italy}
\author{M. Feroci}
\affiliation{\scriptsize INAF/IAPS, via del Fosso del Cavaliere 100, I-00133 Roma (RM), Italy}
\author{A. Ferrari}
\affiliation{\scriptsize CIFS presso Dipartimento di Fisica, Universit\`a di Torino, via P. Giuria 1, I-10125,  Torino, Italy}
\author{F. Fuschino}
\affiliation{\scriptsize INAF/OAS, via Gobetti 101, I-40129 Bologna (BO), Italy}
\author{M. Galli}
\affiliation{\scriptsize INAF/OAS, via Gobetti 101, I-40129 Bologna (BO), Italy}
\affiliation{\scriptsize ENEA Bologna, via don Fiammelli 2, I-40128 Bologna (BO), Italy}
\author{A. Giuliani}
\affiliation{\scriptsize INAF/IASF, via E. Bassini 15, I-20133 Milano (MI), Italy}
\author{C. Labanti}
\affiliation{\scriptsize INAF/OAS, via Gobetti 101, I-40129 Bologna (BO), Italy}
\author{F. Lazzarotto}
\affiliation{\scriptsize INAF -- Osservatorio Astronomico di Padova, vicolo Osservatorio 5, I-35122 Padova (PD), Italy}
\author{P. Lipari}
\affiliation{\scriptsize Dipartimento Fisica, Universit\`a La Sapienza, p.le Aldo Moro 2, I-00185 Roma (RM), Italy}
\affiliation{\scriptsize INFN Sezione di Roma 1, p.le Aldo Moro 2, I-00185 Roma, Italy}
\author{F. Longo}
\affiliation{\scriptsize Dipartimento di Fisica, Universit\`a di Trieste and INFN, via Valerio 2, I-34127 Trieste (TR), Italy}
\author{F. Lucarelli}
\affiliation{\scriptsize SSDC/ASI, via del Politecnico snc, I-00133 Roma (RM), Italy}
\affiliation{\scriptsize INAF -- Osservatorio Astronomico di Roma, via Frascati 33, I-00078 Monte Porzio Catone (RM), Italy}
\author{M. Marisaldi}
\affiliation {\scriptsize Birkeland Centre for Space Science, Department of Physics and Technology, University of Bergen, Norway}
\author{A. Morselli}
\affiliation{\scriptsize INFN Sezione di Roma 2, via della Ricerca Scientifica 1, I-00133 Roma (RM), Italy}
\author{F. Paoletti}
\affiliation{\scriptsize East Windsor RSD, 25A Leshin Lane, Hightstown, NJ-08520, USA}
\affiliation{\scriptsize INAF/IAPS, via del Fosso del Cavaliere 100, I-00133 Roma (RM), Italy}
\author{N. Parmiggiani}
\affiliation{\scriptsize INAF/OAS, via Gobetti 101, I-40129 Bologna (BO), Italy}
\author{A. Pellizzoni}
\affiliation{\scriptsize INAF -- Osservatorio Astronomico di Cagliari, via della Scienza 5, I-09047 Selargius (CA), Italy}
\author{G. Piano}
\affiliation{\scriptsize INAF/IAPS, via del Fosso del Cavaliere 100, I-00133 Roma (RM), Italy}
\author{M. Pilia}
\affiliation{\scriptsize INAF -- Osservatorio Astronomico di Cagliari, via della Scienza 5, I-09047 Selargius (CA), Italy}
\author{C. Pittori}
\affiliation{\scriptsize SSDC/ASI, via del Politecnico snc, I-00133 Roma (RM), Italy}
\affiliation{\scriptsize INAF -- Osservatorio Astronomico di Roma, via Frascati 33, I-00078 Monte Porzio Catone (RM), Italy}
\author{S. Vercellone}
\affiliation{\scriptsize INAF -- Osservatorio Astronomico di Brera, via E. Bianchi 46, I-23807 Merate (LC), Italy}


\shorttitle{\small \agile observations of two repeating FRBs with {\fv low} intrinsic DM}

\shortauthors{\small C.Casentini et al.,}

\begin{abstract}
We focus on two repeating fast radio bursts (FRBs) recently detected by the \chf experiment in 2018--2019 (Source 1: 180916.J0158+65, and Source 2: 181030.J1054+73). These sources have {\mt low excess} dispersion measures (DMs) {\mt ($ < 100 \rm \, pc \, cm^{-3}$ and $ < 20 \rm \, pc \, cm^{-3}$, respectively), implying relatively small maximal distances}. They were repeatedly observed by {\agilep} in the MeV--GeV energy range. We do not detect prompt emission simultaneously with these repeating events. {\fc This search is particularly significant for the submillisecond and millisecond integrations obtainable by \agilep.} The sources are constrained to emit {\mt a} MeV-fluence in the millisecond range below $F'_{MeV} = 10^{-8} \, \rm erg \, cm^{-2}$ corresponding to an isotropic energy {\mt near $E_{MeV,UL} \simeq 2 \times 10^{46}$~erg for a distance of 150 Mpc (applicable to Source 1).} We also searched for $\gamma$-ray emission for time intervals up to 100 days, obtaining {\fvvv 3\,$\sigma$} upper limits (ULs) for the average isotropic luminosity above 50 MeV, {\mt $L_{\gamma,UL} \simeq ${\fvvv \,(5--10)}$\,\times  10^{43} \rm \, erg \, s^{-1}$}. For a source distance {\mt near 100 kpc (possibly applicable to Source 2), our ULs imply $E_{MeV,UL}\simeq10^{40} \rm erg$, and $L_{\gamma,UL} \simeq ${\fvvv \,2}$\,\times 10^{37} \rm \, erg \, s^{-1}$}. Our results are significant in constraining the high-energy emission of underlying sources such as magnetars{\mt, or other phenomena related to extragalactic compact objects,} and show the prompt emission to be lower than the peak of the 2004 magnetar outburst of \sgr {\mt for source distances less than about 100 Mpc}.
\end{abstract}

\section{Introduction}
\label{intro}

Fast radio bursts (FRBs) are a new transient phenomenon of unknown origin consisting of bright millisecond radio pulses (mainly at $\sim 1\, \rm GHz$) having large dispersion measures (DMs) in excess of Galactic values \citep[][]{2007Science.218.777,2019ARA&A119..161101}. For most of the sources with isotropic
 sky distribution, a single radio  burst has been detected. A smaller number of FRBs show repeating pulses that appear to occur erratically in time {\fc \cite[see the FRB catalog, FRBCAT\footnote{http://www.frbcat.org/};][]{2016PASA33..e45}}. No clear evidence for physically different populations has been obtained.

The {\fc possible} cosmological origin of a {\fc sample} of FRBs was confirmed at least for {\fc one source}, the first repeater
 FRB 121102 \cite[][]{2016Natur.531.202S}, which was localized in a dwarf galaxy at z\,=\,0.193 (DM\,=\,560\,$\rm pc\,cm^{-3}$) with a persistent radio emission \cite[][]{2017Natur541...58,2017ApJL834...L8}. Various models have been proposed to explain the FRB radio emission which is characterized by a few-milliseconds timescale and microsecond sub-burst structure \cite[][]{2016Champion,2018Natur.533..132,2019Hessels}. Possible underlying sources include compact objects such as supermassive black holes or young magnetars in young or relatively aged remnants \cite[][]{2019MNRAS.485.4091M}. The enigmatic nature of FRBs results from the milliseconds timescale emission and detected fluences implying energies of radio emission near $10^{39}$\,erg for sources at extragalactic distances. Such large energies and implied radio luminosities suggest a physical production mechanism with no analog found in known Galactic sources.

Recently, the \chf collaboration has reported the discovery of an additional repeating FRB, FRB 180814.J0453 \citep[][]{2019ApJLCHIME,2019ApJLCHIMEb}, and later of eight new repeating sources \citep[][hereafter C19]{2019ApJLCHIMEc} with DM ranging from 103.5 to 1281\,$\rm pc\,cm^{-3}$. The characteristics of this latter sample of repeating FRBs are particularly interesting for counterpart searches, as we elaborate below.
{\fc High-energy counterpart searches for FRBs different from those of {\fc C19} have been recently reported \cite[][]{2017ApJ846...80,2019ApJ879...40,2019SubmittA&A,2020Natur577...190}.}

{\agilep} is a space mission of the Italian Space Agency (ASI) devoted to X-ray and $\gamma$-ray astrophysics \cite[][]{2009A&A...502..995T}, operating since 2007 in an equatorial orbit.
Due to its spinning operational mode, every 7 minutes it exposes $\sim 80\%$ of
the entire sky. The instrument consists of four different  detectors:
an imaging $\gamma$-ray Silicon Tracker {\fc \cite[sensitive in the energy range
30\,$\rm MeV$\,--\,30\,$\rm GeV$;][]{2002NIMPA.490..146B}}, an X-ray imager, Super-\agile {\fc \cite[Super-A,
operating in the energy range 20\,--\,60\,$\rm keV$;][]{2007NIMA581..728}},
the mini-calorimeter \citep[MCAL, sensitive in the
range 0.35\,--\,100\,$\rm MeV$;][]{2008A&A...490.1151M,2008NIMPA.588...17F,2009NIMPA.598..470L}
with $4\pi$ acceptance,
and the anticoincidence (AC) system \cite[][]{2009A&A...502..995T}.
The combination of the Tracker, MCAL, and the AC working together
constitutes the $\gamma$-ray imager (GRID) capable of detecting
$\gamma$-ray transients and gamma-ray burst (GRB)-like phenomena with a good sensitivity.
The GRID has {\mt a field of view (FoV) with a radius of}
70$^{\circ}$ around pointing direction, while Super-A has a {\fc 2D-coded}
 FoV region of radius $\sim\,32^{\circ}$. The MCAL detector is equipped with a
 special triggering system capable of detecting 
MeV transients (e.g., gamma-ray bursts, GRBs) on timescales ranging from sub{\fc millisecond}s to hundreds of seconds.


\agile data are transmitted to the ground at
the ASI Malindi (Kenya) ground station, and delivered to the \agile Data Center
(ADC; part of the ASI Space Science Data Center). Scientific data are then
processed by a fast dedicated pipeline which was recently enhanced for the search of
electromagnetic counterparts of gravitational wave sources \cite[][]{2018IAUS..338...84V,2019ApJ...871...27U,2019RLSFN...33}.
\agile data processing can typically produce  alerts for transient
$\gamma$-ray sources and/or GRB-like events within 20 minutes to 2 hr
from satellite onboard acquisition depending on orbital and
satellite parameters \cite[][]{2013Pittori,2014Bulga}.

We report in this paper high-energy observations by \agile of two repeating FRBs {\fc recently report}ed by CHIME {\fc (C19) which are} characterized by {\fv low intrinsic} DMs (among the smallest ever determined).

\section{Two repeating FRB\small{s} with low intrinsic DM}

Among the sample of {\fc the eight} repeating FRBs of C19, two have quite {\mt low  DMs
in excess of their respective Galactic disk and halo contributions}:
FRB180916.J0158+65 (Source 1), and FRB181030.J1054+73 (Source 2). Table~\ref{tab:tab1} summarizes the characteristics of these repeating radio burst sources, providing the time and the measured radio fluences on millisecond timescales.

 Source 1 is of particular interest because {\fc it} emitted 10 bursts within a 6 month timescale with an average DM\,$\sim 349\,\rm pc\,cm^{-3}$. Its position is near the Galactic plane at coordinates {\fv $l \,=\, 129.7^{\circ}$, $b \,=\, 3.7^{\circ}$}, {\mt within an uncertainty region of 0.09 square degrees}. The range of possible  Galactic contributions to the dispersion measure, DM$_{gal}$, according to two different models \cite[][]{NE2001,YMW16} is between 200 and
 $325\,\rm pc\,cm^{-3}$ {\fc (C19)}.
 {\mt Taking into account the Galactic halo contribution to the DM
 (estimated to be in the range $\rm DM_{halo} \simeq 50-80 \, \rm pc \, cm^{-3}$; \citealt[][]{2019Prochaska})}
  the residual {\mt excess} DM 
  {\mtt is} therefore be quite {\mt low
  ($ \lesssim  100 \, \rm pc \, cm^{-3}$).}
{\mtt The conclusion of C19 was that if Source 1} is extragalactic, it {\mt is} in any case one of the nearest repeating FRBs. Assuming the intergalactic (IGM)-DM relation $DM \simeq 900 \, z \, \rm \, pc \, cm^{-3}$ \cite[][]{2014McQuinn}, we obtain the maximum redshift $ z_{max} \sim 0.1$ for a maximum excess DM of {\mt $100 \, \rm \, pc \, cm^{-3}$ {\fc (C19)}}. {\mtt These considerations have been confirmed by the recent milliarcsecond positioning of Source 1 that appears to be associated with a nearby galaxy at the distance of about 150 Mpc \cite[][]{2020Natur577...190}.}

Source 2, located within an uncertainty region of 0.24 square degrees with centroid Galactic coordinates $l \,=1, 133.4^{\circ}$, $b \,=\, 40.9^{\circ}$, has the lowest overall DM value ever determined {\mt for an FRB} (DM$\,=\, 103.5\, \rm pc\,cm^{-3}$). The range of DM$_{gal}$ for different models is between 32 and  40\,$\rm pc\,cm^{-3}$ {\mt (C19)}.
{\mt Taking into account the Galactic halo contribution $\rm DM_{halo} \simeq 50-80 \, \rm pc \, cm^{-3}$ \cite[][]{2019Prochaska}}, {\mtt the excess DM is near $20 \, \rm pc \, cm^{-3}$ or less;}
the inferred redshift is therefore  $z\,< \,0.023$, which implies a {\mt maximum} luminosity distance less than $\,100\,\rm Mpc$.

\section{\agile Observations}

We searched in the recent \agile archival data for possible exposures of the location regions of Source 1 and Source 2. \agile is operating in spinning mode since 2009 with a revolution of its axis around the Sun-satellite direction every 7 minutes. This mode allows the {\fc \agilep/GRID} to cover 80\% of the sky more than 100 times a day. Gamma-ray exposures of any accessible field last $\sim$\,2 minutes for every satellite revolution with a typical flux sensitivity $\sim 10^{-8}\, \rm erg\, cm^{-2}\, s^{-1}$ above 30\,$\rm MeV$. Furthermore, \agile is the only mission equipped with the onboard submillisecond triggering capability of {\fc \agilep/MCAL}. This makes possible the detection of very fast events in the range 0.4\,--\,100\,$\rm MeV$ (such as Terrestrial Gamma-ray Flashes detected in the submillisecond range; \citealt[][]{2011PhRvL.106a8501T,2014JGRA..119.1337M}).

Based on the FRB data of Table~\ref{tab:tab1}, we first verified the \agile coverage at the times of the FRB occurrences. Depending on data availability, Earth occultation, and detector operations, we first determine for all interesting cases the proper exposure of the \agile detectors (MCAL, GRID and Super-A) at the radio burst times ($T_0$).
It is possible that at $T_0$ the event location is either occulted by the Earth, or it is inside the sky exclusion region in the solar/antisolar directions. We mark such cases with "NO" in Table~\ref{tab:tab1}. We mark with "YES" cases for which the radio event is accessible by the \agile detectors.

\begin{table*}[ht!]
    \caption{Repeating FRBs and \agile observations} 
  \hspace{10mm}
    \begin{tabular}{cclc|ccc} 
     \hline
     \multicolumn{4} {c} {FRB Radio Parameters} & \multicolumn{3} {|c} {Event in the \agile FoV}\\
     \hline
     Label & Day & Arrival Time & Fluence & MCAL & GRID & Super-A\\
     & (yymmdd) & (UTC @ 400\,$\rm MHz$) & (Jy ms) & & & \\
     \hline
     \multicolumn{7} {c} {Source 1: FRB 180916.J0158+65}\\
     \hline
      A & $180916$ & 10:15:19.8021 & 2.3 & NO & NO & NO\\
      B & $181019$ & 08:13:22.7507 & 2.7 & - & - & -\\
      C & $181104$ & 06:57:18.58524 & 2.5 & YES & NO & NO\\
      D & $181104$ & 07:07:01.591 & 2.0 & YES & NO & NO\\
      E & $181120$ & 05:56:06.23243 & 1.8 & YES & NO & NO\\
      F & $181222$ & 03:59:23.2082 & 27 & YES & YES & NO\\
      G & $181223$ & 03:51:28.96040 & 8.1 & YES & YES & NO\\
      H & $181225$ & 03:53:03.9260 & 1.9 & YES & NO & NO\\
      I & $181226$ & 03:43:30.1074 & 3.8 & YES & NO & NO\\
      L & $190126$ & 01:32:45.3289 & 2.0 & - & - & -  \\
     \hline
     \multicolumn{7} {c} {Source 2: FRB 181030.J1054+73}\\
     \hline
     A & $181030$ & 04:13:13.0255 & 7.3 & YES & NO & NO\\
     B & $181030$ & 04:16:21.6546 & 2.2 & YES & NO & NO\\
     \hline
    \end{tabular}
    \label{tab:tab1}
\vskip+.1cm
{\bf Note:}\\
\vskip-.5cm
{\fc\tablenotetext{}{~~ No {\mt \agilep} data are available {\mt for} 181019 and 190126. FRB parameters from C19.}}
\end{table*}
\begin{figure*}[]
\begin{center}
  \centerline{\includegraphics[width=0.5\textwidth]{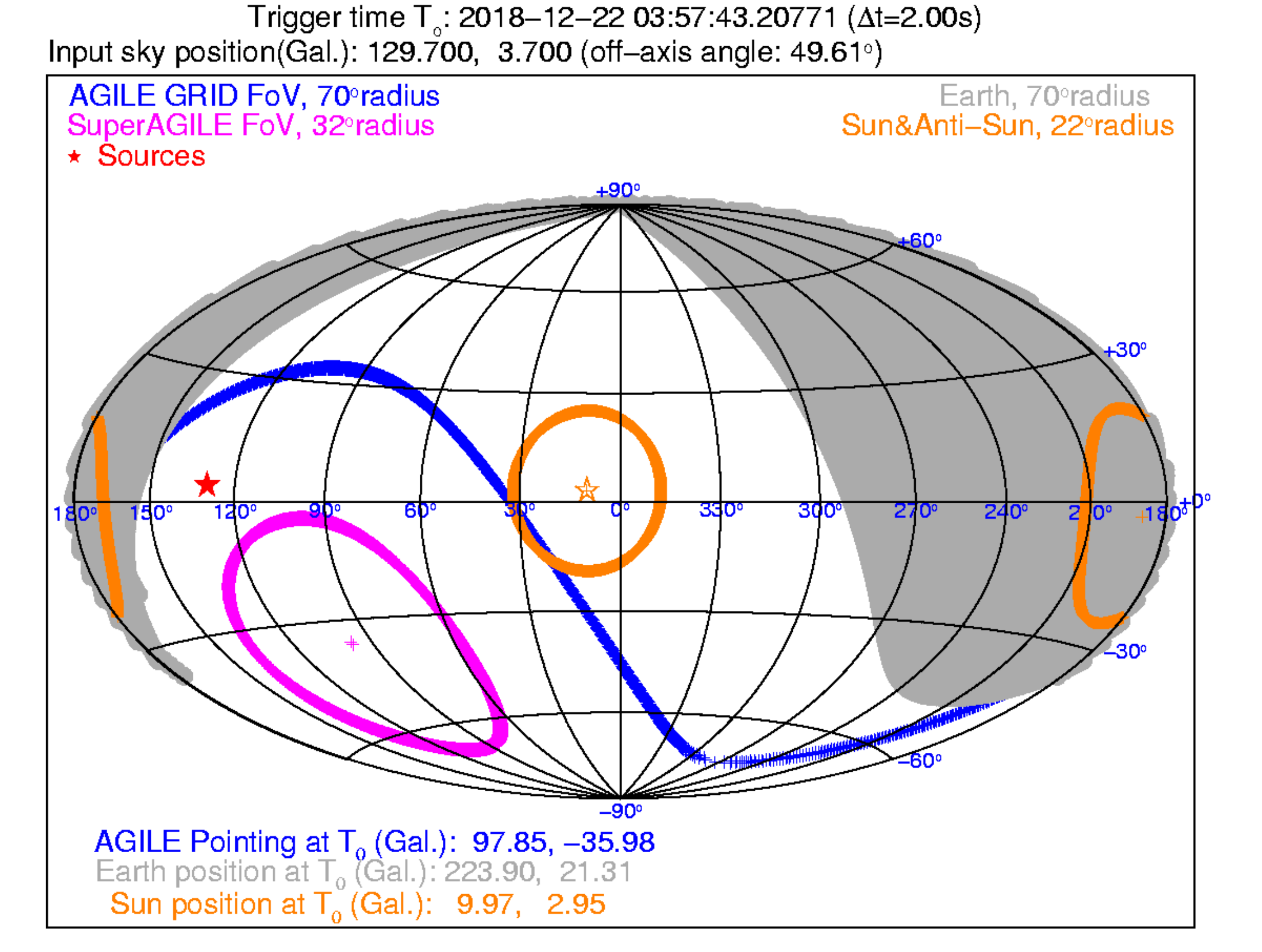}
  \includegraphics[width=0.5\textwidth]{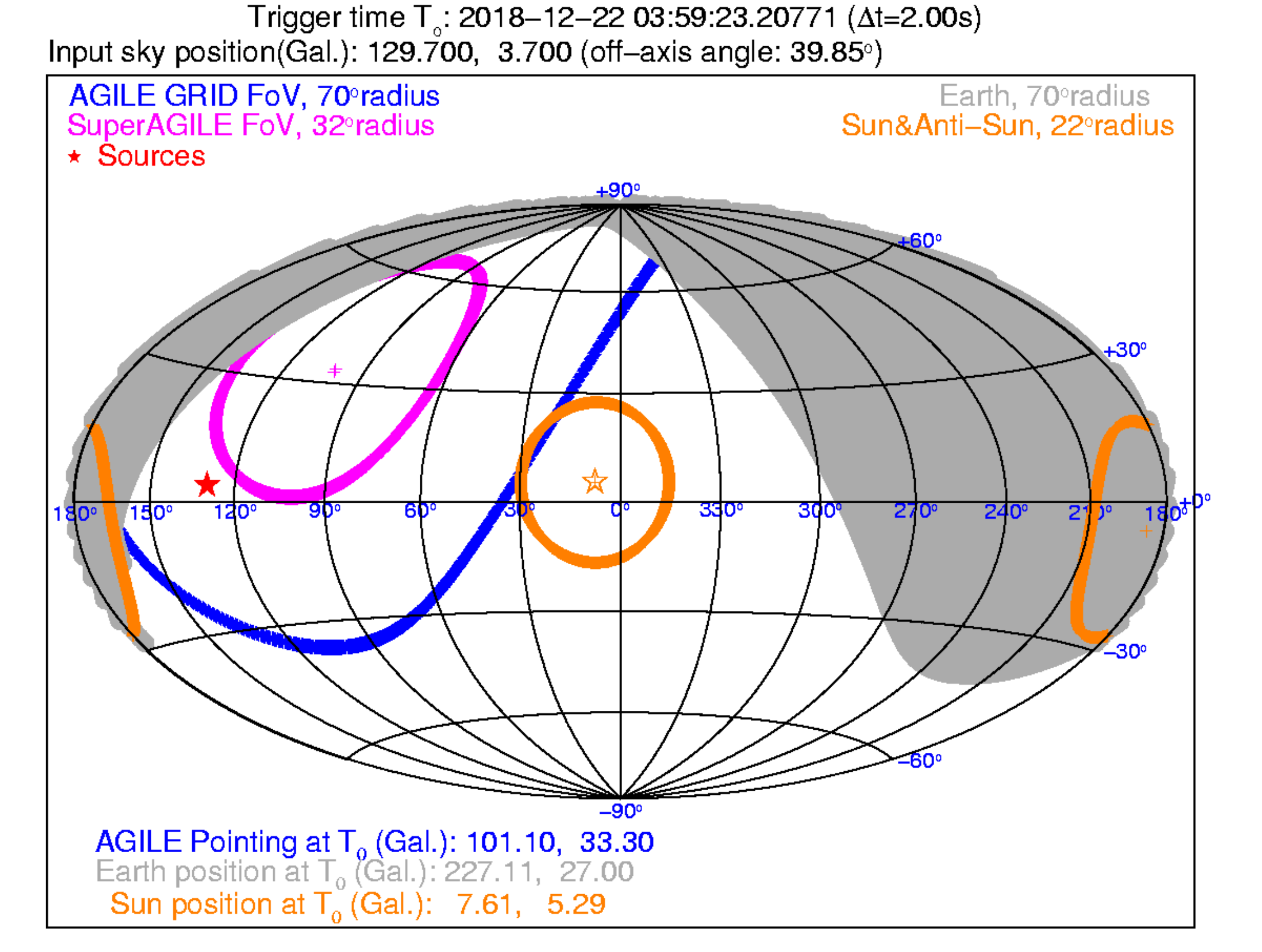}}
  \centerline{\includegraphics[width=0.5\textwidth]{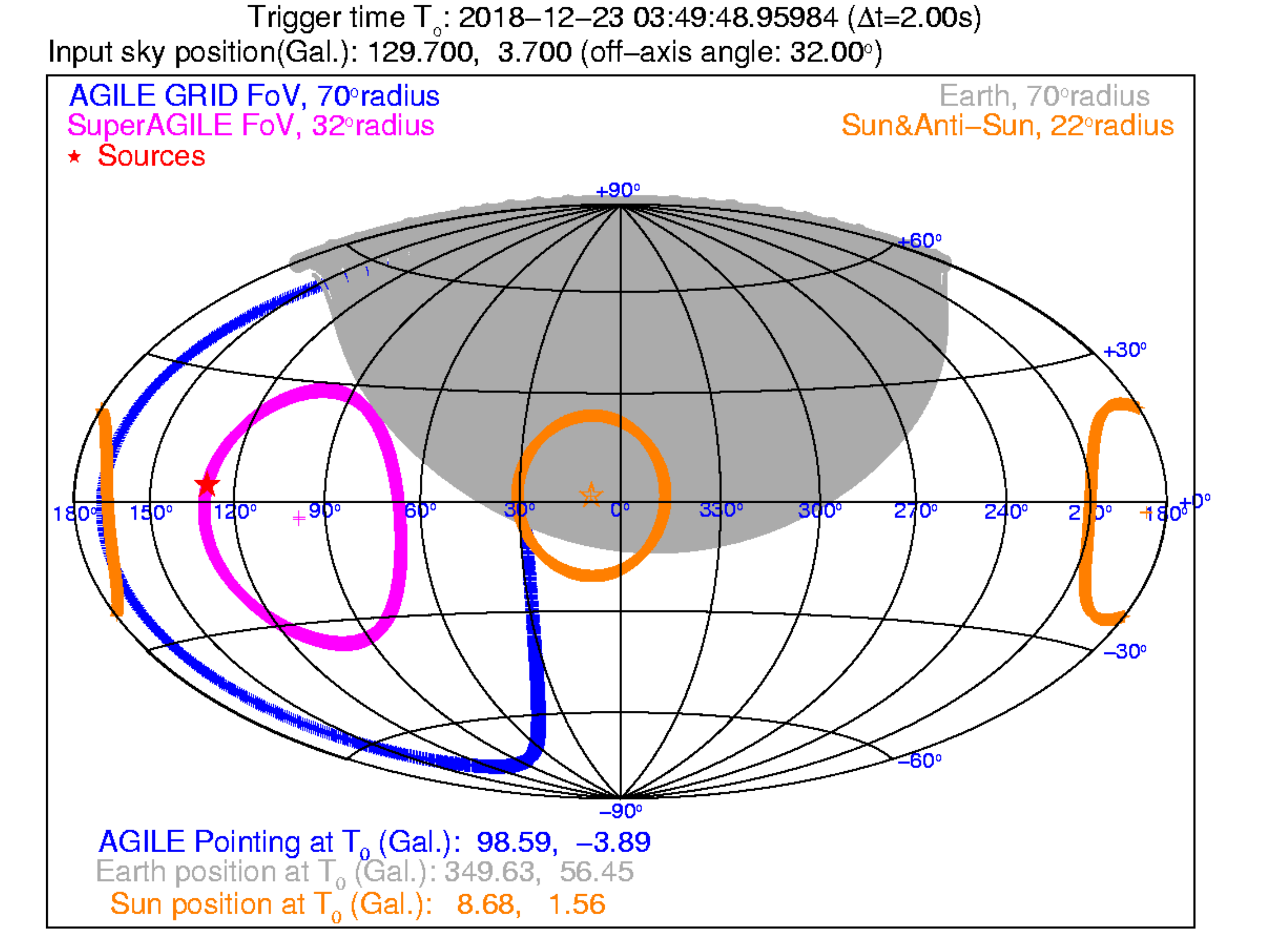}
  \includegraphics[width=0.5\textwidth]{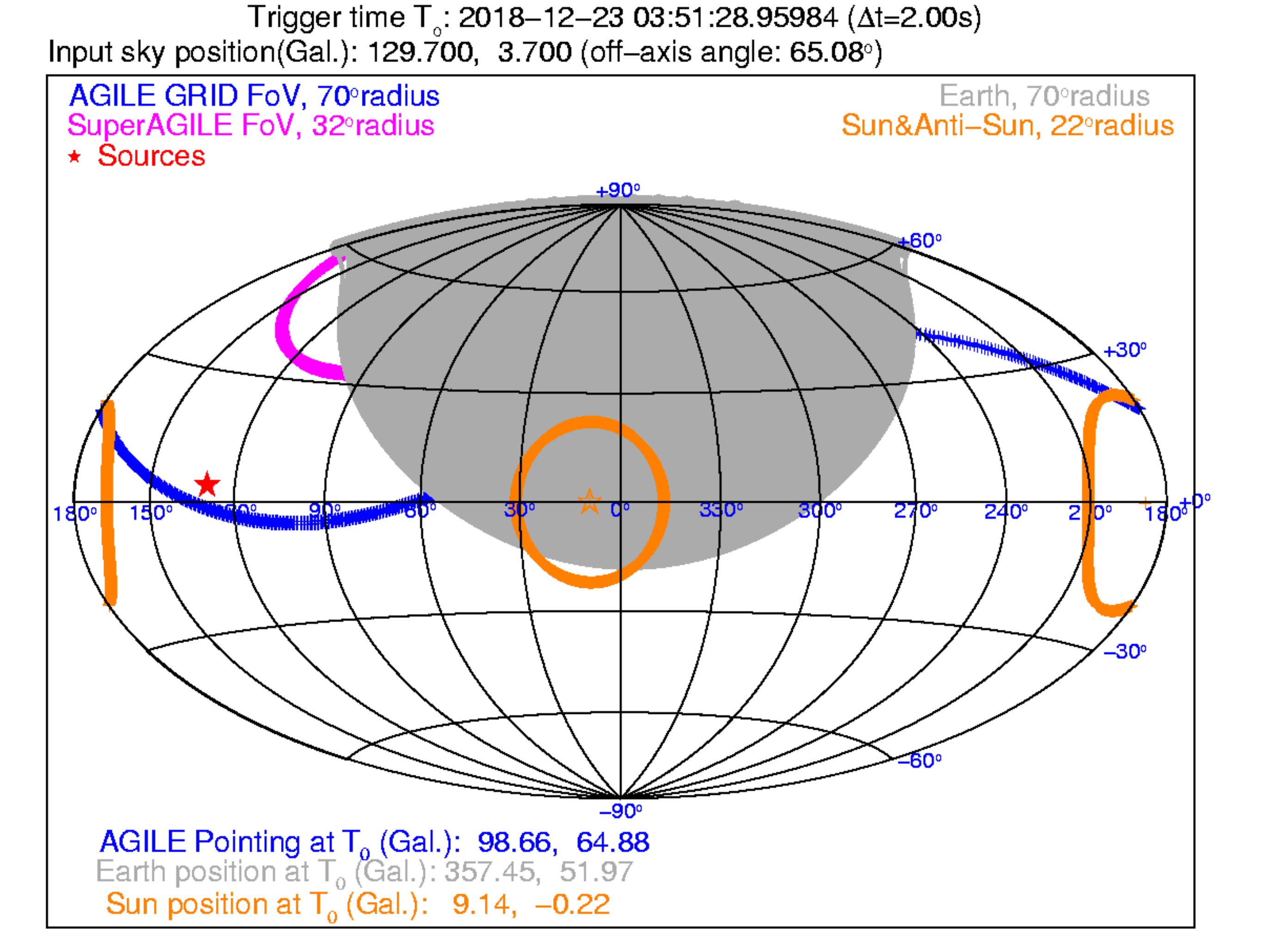}}
  \caption{(\textit{Top panels:}) \agile exposure of burst F of Source 1 (FRB 180916.J0158+65) at ${\fc T_{0}^{1,F}} - 100\, \rm s$ (left panel) and at ${\fc T_{0}^{1,F}}$ (right panel). Plots in Galactic coordinates. The position of the FRB is marked by a red star. The boundaries of the \agilep/GRID and Super-A FoVs are marked in blue and magenta color, respectively.
(\textit{Bottom panels:})
  \agile exposure of burst G of Source 1 at ${\fc T_{0}^{1,G}} - 100 \,\rm s$ (left panel) and at ${\fc T_{0}^{1,G}}$ (right panel). All images are ordered in time from left to right.  }
  \label{fig1}
\end{center}
\end{figure*}

As reported in Table~\ref{tab:tab1}, MCAL had exposure for eight bursts of Source 1; two of them were also in the GRID FoV, and none were in the Super-A one.
Figure~\ref{fig1} (right panels) shows the positions of two events (F and G of Table~\ref{tab:tab1}) of Source 1 and the simultaneous FoV of the GRID and Super-A at their respective $T_0$.

In these plots the FRB positions are marked by red stars, the blue line represents the edge of the GRID FoV, the purple {\fc circle} the edge of Super-A FoV. Every event that falls between the center of the purple circle and the blue line is accessible by the GRID.
In the left panels of Figure~\ref{fig1} we show the GRID and Super-A FoVs at $T_0 - 100\,\rm s$ for which the satellite had {\fc a better} coverage of source position.

\begin{figure*}[]
\begin{center}
  \centerline{\includegraphics[width=0.5\textwidth]{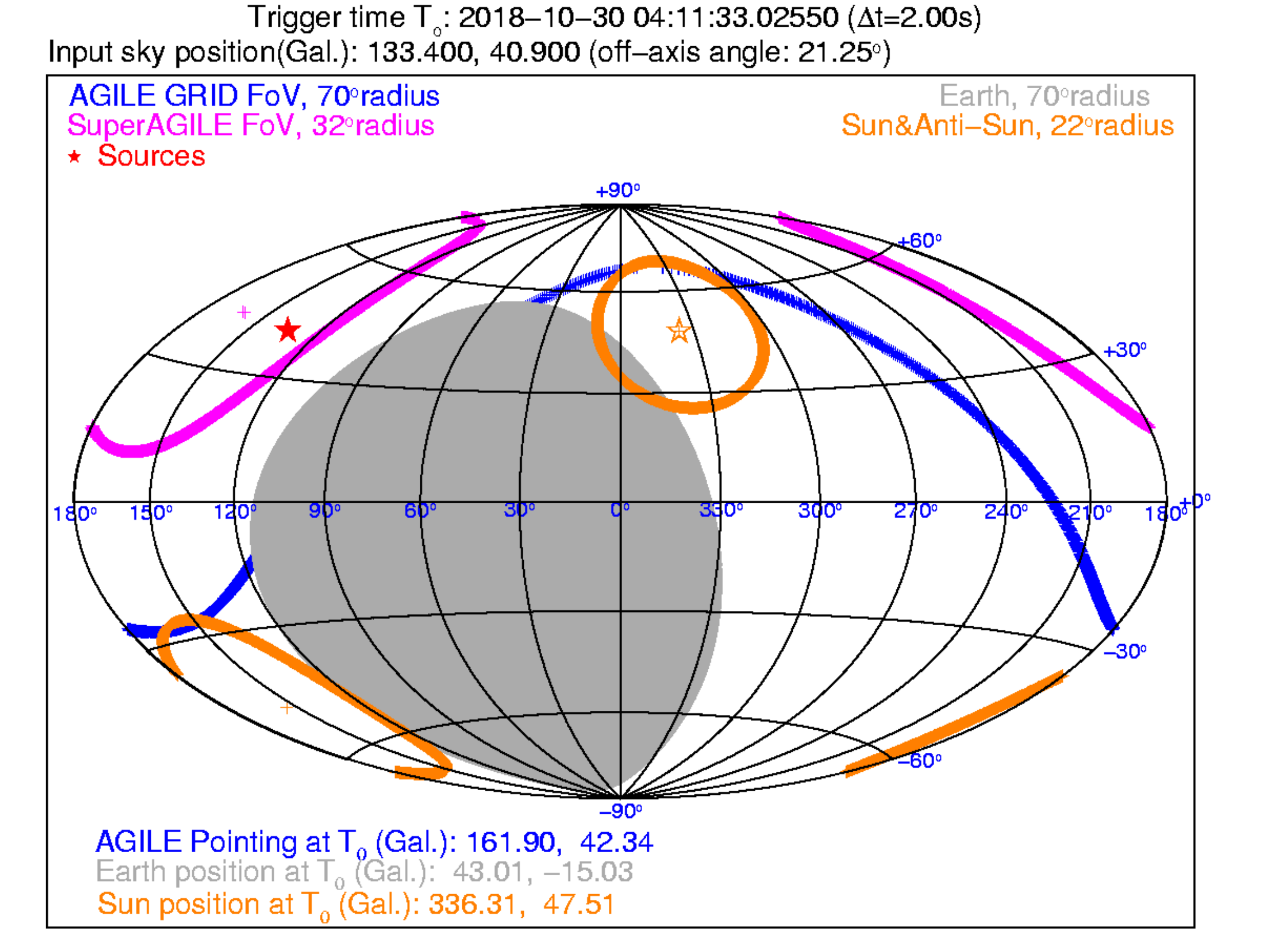}
  \includegraphics[width=0.5\textwidth]{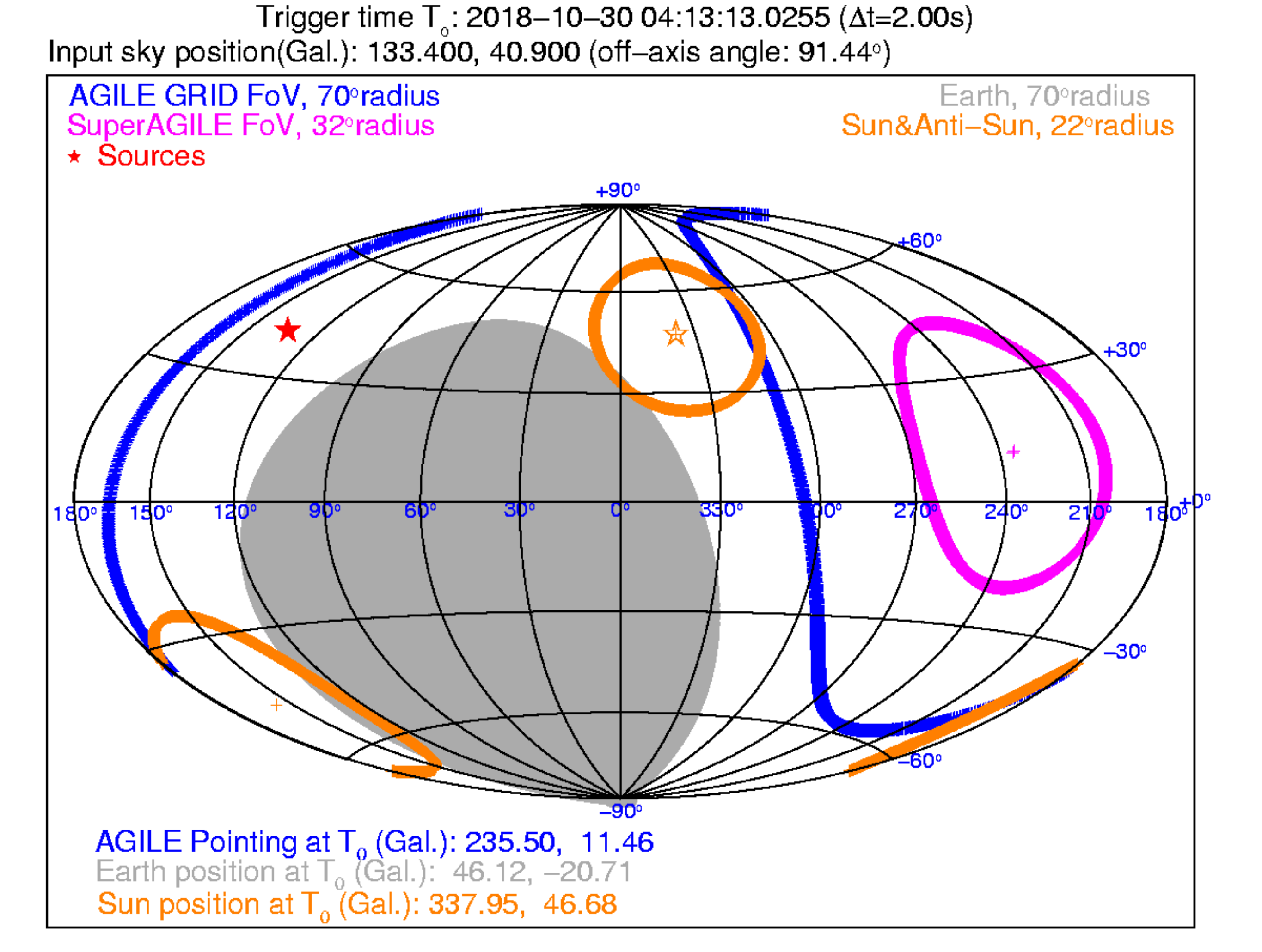}}
  \centerline{\includegraphics[width=0.5\textwidth]{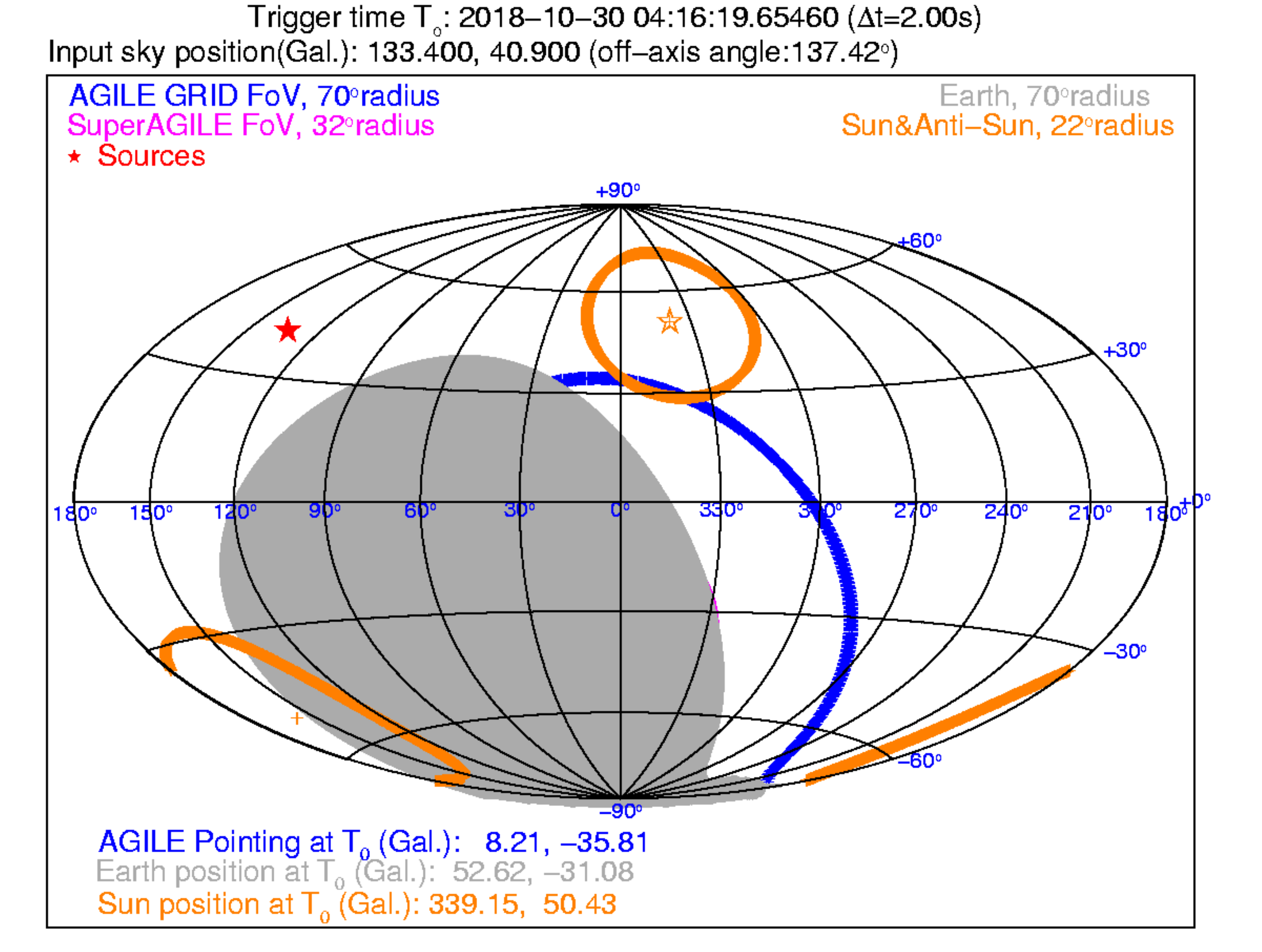}
  \includegraphics[width=0.5\textwidth]{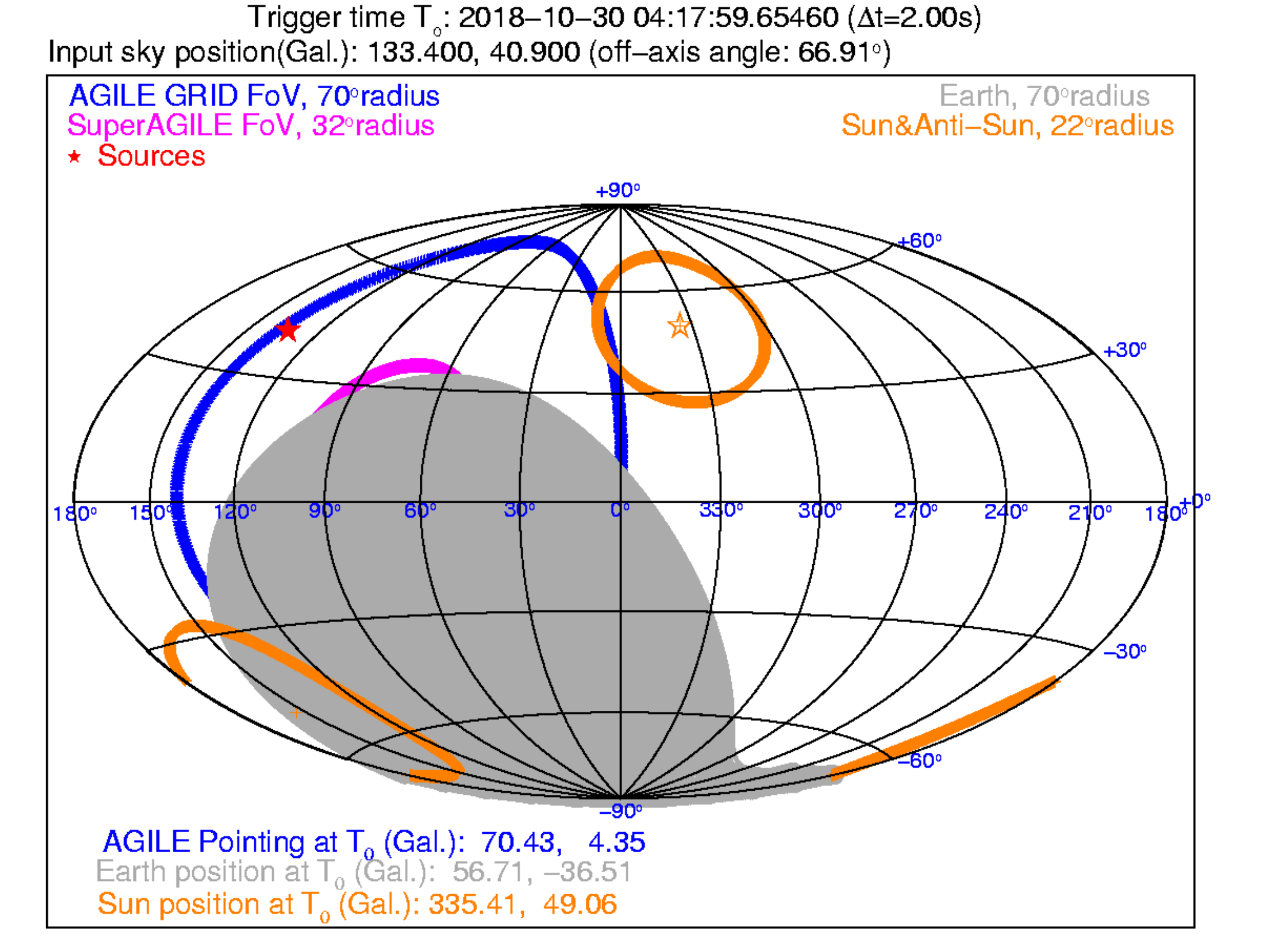}}
  \caption{(\textit{Top panels:}) \agile exposure of burst A of Source 2 (FRB 181030.J1054+73) at ${\fc T_{0}^{2,A}} - 100 \,\rm s$ (left panel) and at ${\fc T_{0}^{2,A}}$ (right panel). Plots in Galactic coordinates. The position of the FRB is marked by a red star. The boundaries of the \agilep/GRID and Super-A FoVs are marked in blue and magenta color, respectively.
(\textit{Bottom panels:})
  \agile exposure of burst B of Source 2 at ${\fc T_{0}^{2,B}}$ (left panel) and at ${\fc T_{0}^{2,B}} + 100 \,\rm s$ (right panel). All images are ordered in time from left to right.}
  \label{fig2}
\end{center}
\end{figure*}

Regarding Source 2, {\fc both bursts were observed by} MCAL but the GRID detector {\fvv never observed the source}
at $T_0$. In Figure~\ref{fig2}, top left panel, is shown the \agile exposure at $T_0 - 100\,\rm s$ (the best available exposure for this radio burst), and in the top right panel the exposure at $T_0$. For the second radio burst of Source 2 in the bottom left panel of Figure~\ref{fig2} is shown the exposure at $T_0$ and in the bottom right panel the best available exposure at $T_0 + 100\,\rm s$. Based on GRID exposures at the $T_0$'s, we focus our subsequent analysis to the bursts F and G of Source 1, and to both bursts (A and B) of Source 2. For all \agile detectors the instruments response depends on the incident angle with respect to the pointing axis as described in e.g., \cite{2009A&A...502..995T,2016ApJ...825L...4T}.

\subsection{MCAL fluence upper limits}

The {\fc \agilep/MCAL} is a detector that operates with seven
different trigger windows, from the 
submillisecond ($\sim\,300\,\mu \rm s$) timescale to a $\sim\,8\, \rm s$ timescale.
{\mtt We searched for impulsive triggered events in the MCAL data for time
intervals centered at the FRB event times $T_0$ provided in Table 1. Details about the onboard trigger logic and data acquisition are given in the Appendix.}

\begin{figure}[]
\begin{center}
  \includegraphics[width=1.\linewidth]{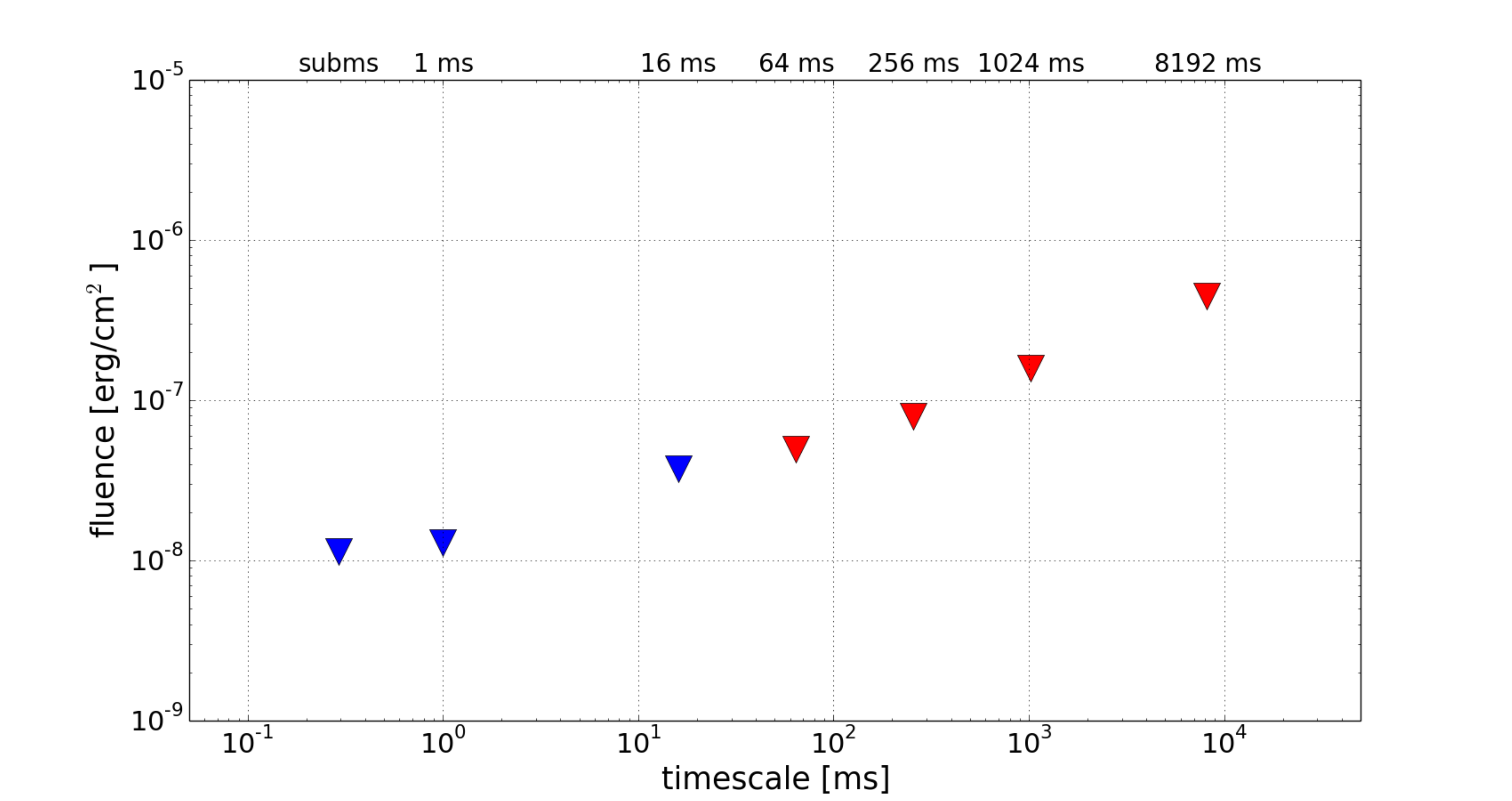}
  \caption{\agilep/MCAL fluence ULs in the energy range 0.4\,--\,{\fvvv 100}\,$\rm MeV$ for the F radio burst of Source 1 as a function of integration timescales{\fc . Similar values are obtained for the burst G}. {\fvv Blue markers show the hardware trigger logic ULs, while red markers show the software trigger logic ones (see the Appendix for details).}}
  \label{fig3}
\end{center}
\end{figure}

{\fc In} Table~\ref{tab:tab1} {\fvv we show} that $70\%$ of Source 1 bursts and $100\%$ of Source 2 bursts are inside the FoV of MCAL. Source 1 burst A is occulted by the Earth and there are no available data at the Source 1 burst B occurrence. No triggers were registered around $T_{0}$'s and no transients were {\fc thus} recorded by MCAL. If MCAL is not triggered, ULs for the fluence can be obtained by evaluating the minimum detectable
flux as a function of angle (based on the detector response matrix) as
 required by the onboard trigger logic at different timescales.
MCAL therefore provides fluence ULs for  the seven different timescales of the
onboard trigger logic, for events that did not trigger the detector. As an example, in Figure~\ref{fig3} are reported the fluence ULs obtained by MCAL for the radio burst F of Source 1. Similar values are obtained {\fc also for burst G} of Source 1 and for those of Source 2.
The {\fc MeV-fluence} ULs obtained for submillisecond and millisecond timescales (being of the order $10^{-8} \, \rm erg \, cm^{-2}$) are particularly {\fc important}, as we discuss below.

\begin{table*}[ht!]
    \caption{\agilep/GRID flux ULs
for the radio burst F, G of Source 1 and burst A of Source 2, as a function of integration timescales.} 
  \begin{center}
    \begin{tabular}{|c|c|c|c|c|c|c|} 
     \hline
     Name & $T_{0}+10^1$ s$^{\tablenotemark{\scriptsize(a)}}$ & $T_{0}+10^2$ s$^{\tablenotemark{\scriptsize(a)}}$ & $T_{0}+10^3$ s$^{\tablenotemark{\scriptsize(a)}}$ & $T_{0}+10^0$ d$^{\tablenotemark{\scriptsize(a)}}$ & $T_{0}+10^1$ d$^{\tablenotemark{\scriptsize(a)}}$& $T_{0}+10^2$ d$^{\tablenotemark{\scriptsize(a)}}$\\
     \hline

     Source 1-F & 4.0$\times 10^{-7}$ & 3.6$\times 10^{-8}$ & 1.5$\times 10^{-8}$ & 4.1$\times 10^{-10}$ & 1.1$\times 10^{-10}$ & 3.9$\times 10^{-11}$\\  
     Source 1-G & 1.7$\times 10^{-6}$ & 3.4$\times 10^{-8}$ & 1.6$\times 10^{-8}$ & 4.1$\times 10^{-10}$ & 1.1$\times 10^{-10}$ & 4.0$\times 10^{-11}$\\
     Source 2-A & 2.7$\times 10^{-7}\,{\fc \tablenotemark{\scriptsize(b)}}$ & 4.7$\times 10^{-8}\,{\fc \tablenotemark{\scriptsize(b)}}$ & 1.0$\times 10^{-8}$ & 7.3$\times 10^{-10}$ & 5.3$\times 10^{-11}$ & 1.8$\times 10^{-11}$\\

     \hline
    \end{tabular}
    \label{tab:tab2}
 \end{center}
\vskip-.1cm
{\fc {\bf Notes:}\\
\vskip-.4cm
{(a):}{ {\fvvv 3}\,$\sigma$ flux ULs ($\rm erg \, cm^{-2} \, s^{-1}$) obtained for emission in the range 50\,MeV\,--\,10 GeV for the short inte\-gration timescales and 100\,MeV\,--\,10 GeV for the long ones, at each source positions.}\\
{(b):}{~~The first two ULs of Source 2 are evaluated on intervals [+390, +400]\,$\rm s$ and [+390, +490]\,$\rm s$, respectively. These two intervals {\fv start} as the source{ \fv enters the exposed region} during the {\mt \agilep} spinning {\mt revolution}.}}
\end{table*}

\subsection {GRID observations  of Source 1 (FRB 180916.J0158+65)}

{\fc As reported in Table~\ref{tab:tab1}, GRID} has exposure of burst F and G of Source 1 at $T_0$.
The radio burst positions are inside the GRID FoV as shown in the right panels of Figure~\ref{fig1}.

There is no known $\gamma$-ray source within a circle of radius 10 degrees, the usual \agile {\fv M}aximum {\fv L}ikelihood region of interest, at the position of the FRB. Since \agile rotates every $\sim\,7$ minutes around a direction pointing at the Sun, the GRID obtained exposures of the source for each satellite revolution not affected by Earth occultation and/or SAA passages. We consider two types of
{\fc analysis \cite[as in][]{2017ApJ...850L...27V}: (1) the "GRB detection mode" for short integration times \cite[see][]{2010ApJ...708L..84G} within the interval [0, +1000] s (times refer to $T_0$) from 10 to 1000 s integration times, in the 50\,$\rm MeV$\,--\,10\,$\rm GeV$ band; (2) the \agile Maximum Likelihood analysis \cite[][]{2012Bulgarelli} for long integration times, ranging from 1 day to 100 days, in the 100\,$\rm MeV$\,--\,10\,$\rm GeV$ band}.

{\fc The short exposures of Source 1, with 10, 100 and 1000 s integration times during the first passage on its position, allowed us to obtain the {\fvvv 3}\,$\sigma$ flux ULs in the 50\,$\rm MeV$\,--\,10\,$\rm GeV$ band reported in Table~\ref{tab:tab2}; they range from $1.7 \times 10^{-6}\,\rm erg\, cm^{-2}\, s^{-1}$ to $1.5\, \times \,10^{-8}\,\rm erg\, cm^{-2}\, s^{-1}$. The UL values of bursts F and G are compatible}, with the exception of UL$_{10s}^{1,F}$ that is $\sim 4$ times smaller with respect to UL$_{10s}^{1,G}$: this is because at $T_{0}$ the source was at the edge of the GRID FoV {\fc which is moving away from the source region}. Its effective exposure is lower than in the same interval for Burst F.

\begin{figure*}[]
\begin{center}
   \centerline{
  \includegraphics[width=0.6\linewidth]{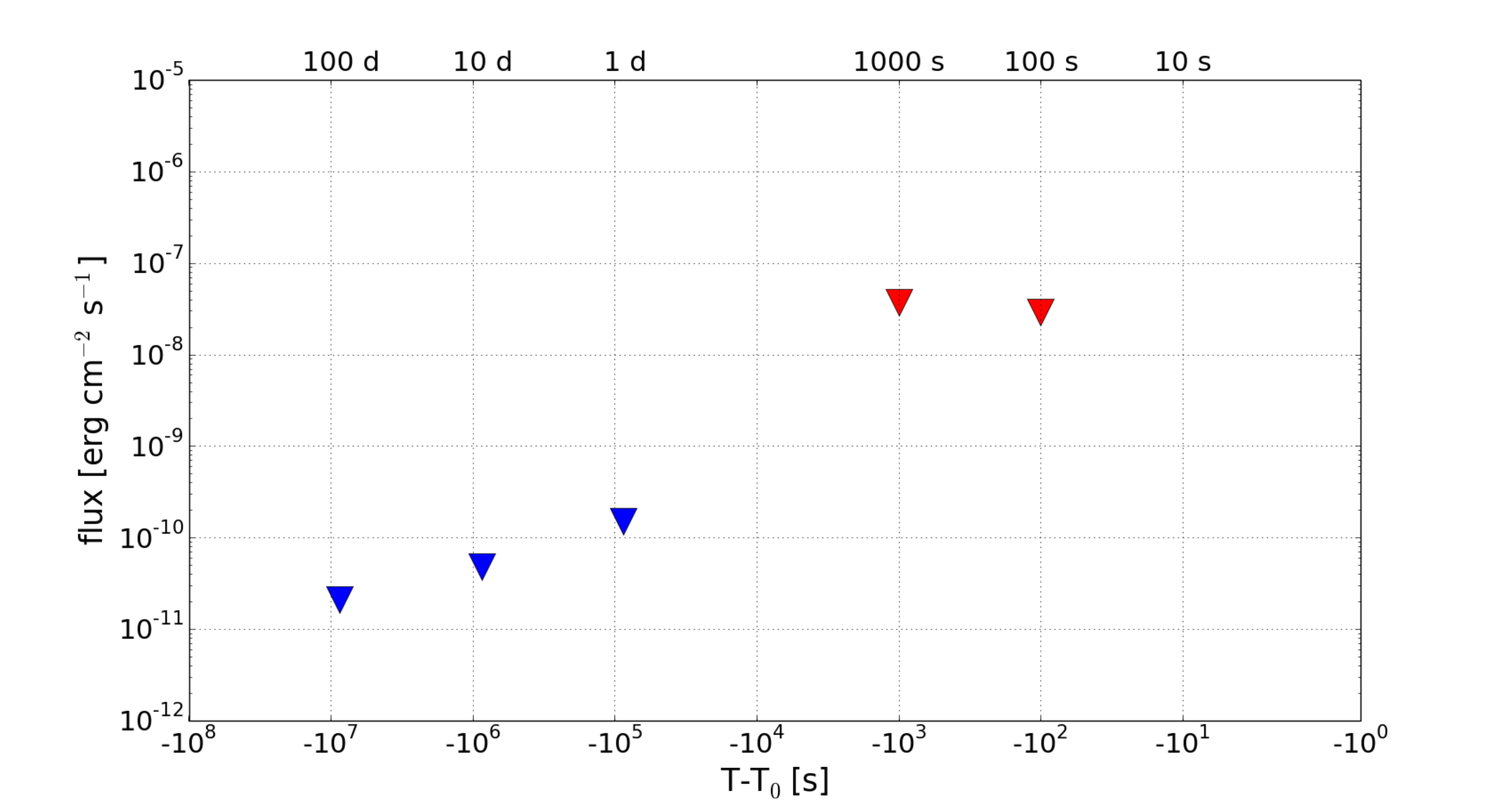}
 \hspace{-1.2cm}
  \includegraphics[width=0.6\linewidth]{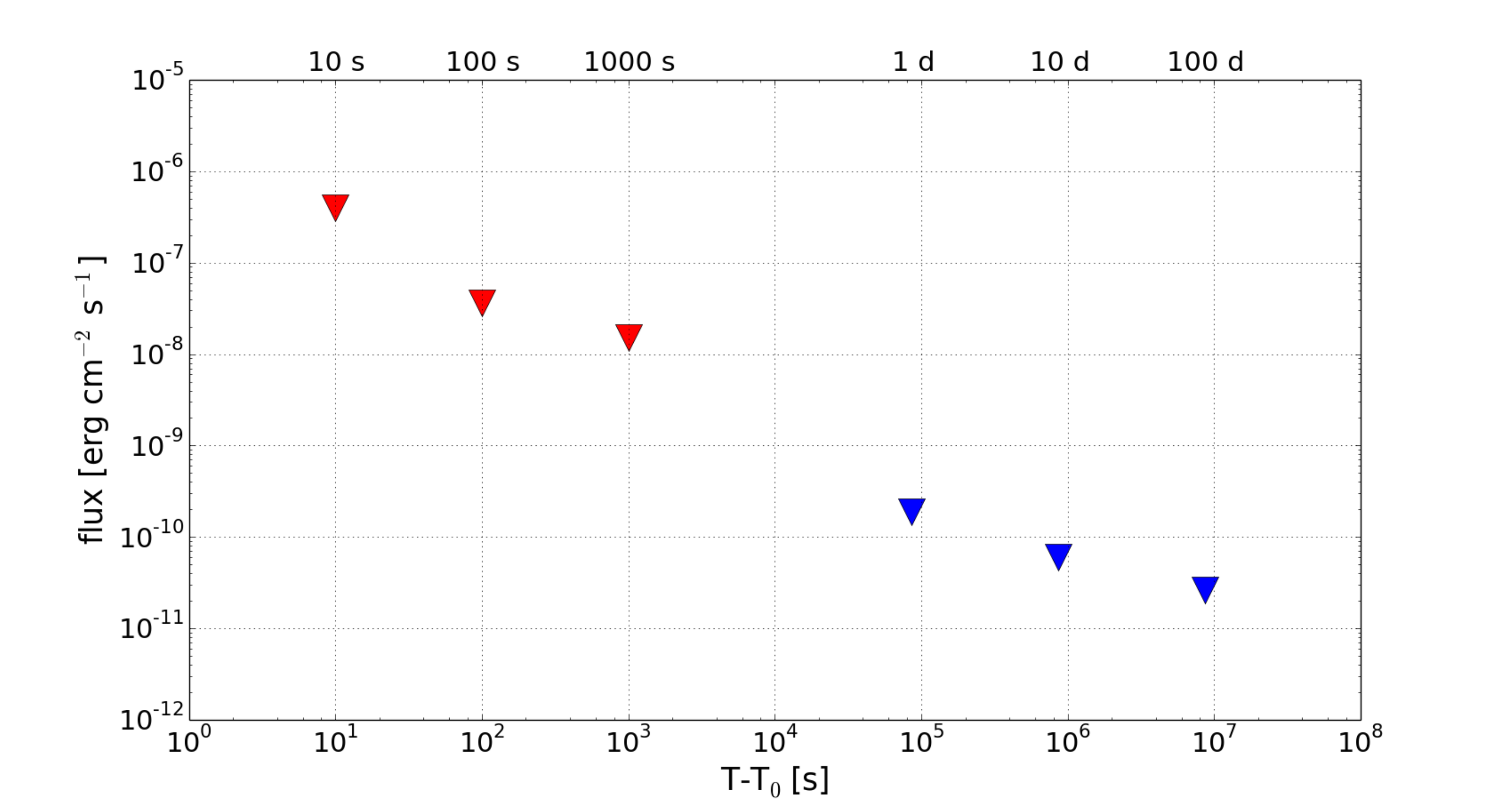}
   }
  \caption{\agilep/GRID {\fc {\fvvv 2\,$\sigma$} flux} ULs for the F burst of Source 1 as a function of {\fc integration} times{ \fvv pre- and post-burst $T_0$. The symbols mark the beginning (left panel) or the ending (right panel) of the integration time intervals that have FRB $T_0$ as one extreme time value. The short timescale values depend on the specific exposure and pointing geometry of the \agile satellite}. {\fc The energy ranges are 50\,$\rm MeV\, < E <\,10\, \rm GeV$ for timescales 10, 10$^2$, 10$^3 \rm s$ (red markers), and 100\,$\rm MeV\, < E <\,10 \,\rm GeV$ for longer integrations (blue markers).}}
  \label{fig4}
\end{center}
\end{figure*}

{\fc The long exposures (with integrations of 1, 10 and 100 days) lead to the {\fvvv 3}\,$\sigma$ flux ULs (see Table~\ref{tab:tab2}) in the 100\,$\rm MeV$\,--\,10\,$\rm GeV$ band ranging from {\fvvv 4.1}$\times 10^{-10}$ erg cm$^{-2}$ s$^{-1}$ to {\fvvv 3.9}$\, \times  10^{-11}$ erg cm$^{-2}$ s$^{-1}$.}
In Figure~\ref{fig4}, we {\fvv show} our $\gamma$-ray ULs for Source 1 burst F {\fvv that we obtained for time intervals before and after the event. The short timescale values depend on the specific exposure and pointing geometry of the \agile satellite.}
\subsection {GRID observations of Source 2 (FRB 181030.J1054+73)}

None of the two bursts of Source 2 are inside \agilep/GRID FoV at $T_{0}$, as {\fc shown} in {\fc the} top right panel and {\fc the} bottom left panel of Figure~\ref{fig2}. Also {\fc in this case}, we split the analysis in short and long timescales, using the same integration times as in the previous case. Furthermore, due to the small amount of time between the two bursts ($\sim $ few minutes), we made the analysis considering only the first repetition.
{\fc The short exposures allowed us for this source to obtain the {\fvvv 3}\,$\sigma$ flux ULs in the 50\,$\rm MeV$\,--\,10\,$\rm GeV$ band reported in Table~\ref{tab:tab2}), ranging from $2.7\,\times 10^{-7}$ erg cm$^{-2}$ s$^{-1}$ to $1.0 \times 10^{-8}\, \rm erg\, cm^{-2}\, s^{-1}$.
The 10 and 100 s integration ULs were evaluated on intervals [+390, +400]\,$\rm s$ and [+390, +490]\,$\rm s$, respectively, when the source entered the GRID FoV. For longer exposures we obtain ULs ranging from $7.3 \times 10^{-10}\,\rm erg\, cm^{-2}\, s^{-1}$ to $1.8 \times 10^{-11}\,\rm erg\, cm^{-2}\, s^{-1}$, in the 100\,$\rm MeV$\,--\,10\,$\rm GeV$ energy band (see Table~\ref{tab:tab2})}.

\newpage
\section{Discussion}

The energy of the FRB radio burst emission is believed to be a small fraction of the
total energy involved in any plasma configuration capable of radiating isotropically-estimated energies
of $10^{39} \rm erg$ or larger for FRBs with large DMs \cite[e.g.,][]{2019ARA&A119..161101}. If the large-DM FRBs are indeed at extragalactic distances, the underlying energy sources are most likely associated with
phenomena involving compact objects, either neutron stars or black holes in peculiar states of emission.
Therefore, constraining the energetics of the underlying plasma source for small-DM FRBs offers a unique way of exploring energy and luminosity ranges that are substantially smaller than those applicable to large-DM/extragalactic FRBs.{ \fc Observed quantities depend on unknown beaming. In the
following we assume isotropic emission for our estimates of energies and luminosities.}
Our {\mtt reference} fluence $F'$ in the millisecond range is the typical \agilep/MCAL fluence UL
obtained for several events of Table~\ref{tab:tab1}, that is $F'_{MeV} = 10^{-8} \, \rm erg \, cm^{-2}$.

{\mt Let us consider first the extragalactic distance hypothesis for both Source 1 and Source 2.} {\mtt We assume here the distance $d = 150$ Mpc for both Source 1 and Source 2, in agreement with the recent detection of Source 1 by \cite{2020Natur577...190}.} Our value of $F'$ translates into an UL for the radiated energy into the MeV range,
{\mt $ E_{MeV,UL} =\, 4 \, \pi F' \, d_{150M}^2 \simeq (2.2 \,\times  10^{46} \, \rm erg) \, d_{150M}^2$, where $\rm d_{150M} = \, d/150\, \rm Mpc$ } with $d$  the source distance from Earth.
This value of $ E_{MeV,UL} $ should be compared with the energy involved in the radio burst observation,
{\mt $E_{radio, iso} \simeq  (2.7\,\times\, 10^{37} {\rm erg}) \, S_{\nu, Jy} \, \delta t_{ms} \,
\Delta \nu_{GHz} \, d_{150M}^2$}, where the measured FRB flux density in the radio band is $S_{\nu,Jy}$ in units of jansky,
$\delta t_{ms}$ is the temporal width in units of milliseconds, and $\Delta \nu_{GHz}$ is the radio bandwidth in
units of GHz. Note that events F and G of Source 1 have $ S_{\nu, Jy} \sim 10$. If the two energies are compared for a time interval (e.g., milliseconds) assumed to be the same for the radio and MeV emissions, the ratio $E_{radio, iso}/{\fc E_{MeV,UL} \gtrsim\,} 10^{-8}$ clearly indicates
the marginal contribution of the radio emission from the point of view of the overall energetics for events that are expected to be detected in the MeV range.
MCAL UL's for 1\,-\,10 s integrations are larger by more than one order of magnitude compared with that of millisecond timescales (see Figure~\ref{fig3}), and the corresponding UL on the involved energy is {\mt $ E_{MeV,UL} \simeq (\rm a \, few  \, 10^{47} \,  erg) \, d_{150M}^2$}.

{\mtt If Source 2 is located within our Galaxy or Galactic halo, our results are  more constraining}. In this case,
our value of $F'$ translates into the UL,
 $ E_{MeV,UL} =\, 4 \, \pi F' \, d_{100k}^2 \simeq (10^{40} \, \rm erg) \, d_{100k}^2$, where $\rm d_{100k} =~d/100\, \rm kpc$, and the energy involved in the radio burst is
{\mt $E_{radio, iso} \simeq  (1.2\,\times 10^{31} {\rm erg}) \, S_{\nu, Jy} \, \delta t_{ms} \,
\Delta \nu_{GHz} \, d_{100k}^2$}.

MeV-radiated energies of order of $10^{40} - 10^{41}\,\rm erg$ are
 {\mt relevant} for systems hosting a magnetar-like star, that is for {\mt compact
  objects} whose outbursts are energized by either magnetic instabilities or by magnetospheric acceleration phenomena. The very large magnetic fields associated with magnetars (of the order of $B_m \sim 10^{14} - 10^{16} \, \rm G$) in principle can be associated with maximal energies $E_m \sim R_m^3 B_m^2 / 6 \simeq (2\,\times 10^{49} erg)  \, B{\fc ^{2}}_{m,16}$, with $B_{m,16}$ the magnetar inner magnetic field in units of $10^{16}\,\rm G$ and where we assumed a magnetar radius $R_m = 10^6 \, \rm cm$.
It is not clear how and when fractions of the total magnetic energy can be dissipated 
 {\mt by} special instabilities. The analysis of \citealt[][]{thompson1996} implies a timescale of $\tau_a \sim 10^{11} \, \rm s$ for ambipolar diffusion of the inner magnetar field, and therefore an average luminosity
\be L_a \sim E_m / \tau_a \sim (10^{38} \, {\rm erg \,s^{-1}}) \, E_{m,49} \, \tau_{a,11}^{-1}
\label{eq-1}  \en
where $E_{m,49}$ is the maximal energy in units of $10^{49} \, \rm erg$, and
$\tau_{a,11}$ is the diffusion time in units of $10^{11} \, s$. Smaller values of $\tau_a$ \cite[e.g.][]{beloborodov17} {\fc and/or of $B^{2}_{m,16}$ imply} larger{\fc /smaller} average luminosities. We consider the luminosity $L_a$ of Eq. \ref{eq-1} as a reasonable reference for comparison with our observations.

\agile short time observations summarized in Table~\ref{tab:tab1} and Figs. 1-3 
 constrain any flaring activity of a putative magnetar {\mtt associated with Source 2 and} subject to sudden instabilities. Any MeV-radiated energy associated with simultaneous FRB activity is constrained to be less than $ E_{MeV,UL} \sim (10^{40} - 10^{41} \rm erg) \, d_{100k}^2$.
  {\mtt Note that the former value applies for timescales of milliseconds or less, and the latter value to timescales near 200 ms}.
  When compared with the 2004 powerful flare of {\sgr}that radiated $E_{MeV,SGR} \simeq 2 \,\times 10^{46} \, \rm erg$ {\mtt within $\sim 200$ ms} in the MeV range (\citealt[][]{palmer2005}; {\mtt \citealt[][]{2005Natur434...1098}}) our observations clearly exclude such an occurrence. If we use the observed {\mtt peak energy} of $10^{46}\,\rm erg$ in obtaining the fraction $\xi$ of available magnetic energy being radiated into the MeV range, we deduce $\xi \sim E_{MeV,SGR}/E_m \sim 10^{-3} \, B_{16}^{-2}$. We therefore obtain the ULs of the flaring energies {\mtt associable} with the repeating FRBs of Table~\ref{tab:tab1}, $E_* \sim E_{MeV,UL}/\xi \sim (10^{43}\, -\, 10^{44}\, \rm erg) \, B_{16}^{-2} \, d_{100k}^2$. These values exclude major magnetic flaring activity from magnetars with
inner magnetic fields near $10^{16}\,\rm G$.
Therefore, either the magnetic field of the associated magnetar is
smaller than $10^{16}\,\rm G$, or the {\mtt intensities} of the MeV flares {\mtt that we are trying to detect} are substantially smaller than those observed in the case {\mtt of the giant flare of} \sgrp. Both of these possibilities are reasonable, and we consider these constraints an interesting outcome of our observations.

{\mtt Additional constraints derive from \agile gamma-ray observations
of Source 1 and Source 2 with long time integrations.}
\agilep/GRID 
{\mtt measurements at} the locations of the repeating
FRBs of Table~\ref{tab:tab1} constrain 
possible high-energy
emission above 50 MeV on intermediate and long timescales as summarized
in Table~\ref{tab:tab2}. If we consider the long timescale flux ULs for
100 day integrations, we obtain
$F_{\gamma} \sim \,${\fvvv (2--4)}$\,\times 10^{-11} \rm \, erg \, cm^{-2} \, s^{-1}$.
Therefore, for {\mtt the 150 Mpc distance of Source 1 we
obtain an UL on the isotropic luminosity emitted above 50 MeV,
$L_{\gamma,UL} \simeq \, (${\fvvv 5--10}$)\,\times 10^{43} \rm \, d_{150Mpc}^2 \, erg \, s^{-1}$}.
{\mtt For an assumed distance within our Galactic halo of Source 2},
we obtain
\be   L_{\gamma,UL} \simeq {\fvvv 2} \,\times 10^{37} \rm \, d_{100k}^2 \, erg \, s^{-1}.  \label{eq-2} \en
This luminosity should be compared with that of Eq. \ref{eq-1}. A comparison of these two luminosities of the same order of magnitude suggests a not very constraining UL on the 50 MeV\,-\,efficiency $\xi_{GeV} \lesssim \fvvv 0.2$. On the other hand, ULs of the type of Eq. \ref{eq-2} are the lowest ever obtained in the $\gamma$-ray energy range {\mt for a possibly {\mtt FRB} nearby source}, and constrain the flaring activity of the underlying {\mt object} {\mtt of Source 2 (that may turn out to be closer than Source 1 based on its excess DM)}.

\section{Conclusions}
\label{concl}

Fast radio bursts continue to be puzzling with features that are not satisfactorily explained in current modelling of compact objects. A crucial diagnostic of the physical processes involved in FRBs is provided by X-ray and $\gamma$-ray simultaneous observations. In this paper we focused on two distinct FRB sources that erratically repeat and that have {\fvv low intrinsic DM} {\mtt indicating  relatively small distances}. From the hypothesis that the {\mtt excess} DM is directly related to distances, these two FRBs can be considered among the closest ever detected in the radio band. They are therefore the most interesting for constraining their high energy emission. {\mtt The recent detection of Source 1 associated with a nearby spiral galaxy at 150 Mpc distance \cite[][]{2020Natur577...190} adds weight to this prospect.
It is interesting to note that if also Source 2 is located outside our Galaxy,
a UL on its distance is of the order of 100 Mpc (as noticed by C19).}


\agile observations provide important constraints on the prompt
{\mtt (millisecond and hundreds of millisecond timescales)} emission
in the sub-MeV/MeV range, and exclude strong magnetar flares
of the type of the 2004 event from \sgrp. This is especially true for
{\mtt a Source 2 Galactic halo distance}, but applies also to distances
{\mtt of the order of} 100 Mpc. Our UL of the MeV isotropic emitted energy
at {\mtt millisecond/hundreds of millisecond} timescales
{\mt $E_{MeV,UL} \sim (10^{40} - 10^{41} \rm \, erg) \, d_{100k}^2$}
constrains magnetar models {\mtt for Source 2}. For larger distances,
{\mtt our} energy UL is less compelling, but in any case reaches values
of the order of 
the emitted MeV energy from the giant 2004 outburst of \sgr
{\mtt during its first few hundreds milliseconds of emission}.
{\mtt Indeed, we note}
that the maximum distance out of which we can exclude such a large flare from {\mtt FRB sources} is of the order of 100 Mpc.

Furthermore, we constrain the persistent long timescale $\gamma$-ray emission above 30 MeV from the repeating FRBs of Table~\ref{tab:tab1}.
{\mtt For locations up to 150 Mpc, our ULs constrain the $\gamma$-ray emitted power to be  smaller  than $L_{\gamma,UL} \simeq 5 \,\times 10^{43} \rm \, d_{150M}^2 \, erg \, s^{-1}$, i.e., substantially smaller than those detected by nearby blazars.}
{\mtt If the Source 2 location is within our Galactic halo,} our UL of the average isotropic $\gamma$-ray luminosity, $L_{\gamma,UL} \simeq 2 \,\times 10^{37} \rm \, d_{100k}^2 \, erg \, s^{-1}$,
excludes steady {\mtt magnetar-like} power to $\gamma$-ray conversions of the order of 10\% for the source in the Galactic halo.

In conclusion, \agile observations constrain the activity of the underlying energy source both on short and long timescales. Major flaring activities of the type of {\sgr} are clearly excluded {\mtt for a Source 2 within our Galactic halo}, with the \agile MeV fluence ULs reaching values nearly 3 orders of magnitude less {\mt than the {\sgr} flare} for {\mtt a nearby source}. The $\gamma$-ray limits on the isotropic luminosity obtained above 30 MeV are somewhat less constraining. In any case, one can exclude the presence of an energy source at levels reported for different timescales in Table~\ref{tab:tab2}.

FRBs continue to be sources of great interest. More observations are needed to further constrain the emission process and the physical properties associated with the detected radio bursts. {\mtt Having determined the distance of Source 1 is of great relevance. Establishing the distance of Source 2 (which has the smallest excess DM ever measured)  would also be invaluable in constraining models.} \agile continues its observations of the $\gamma$-ray sky, and will provide useful data for the search of FRB high-energy counterparts.
\\
\\
\\

{\bf Acknowledgments:}  We acknowledge comments by {\mtt two anonymous referees} that
were taken into account {\mtt  and contributed in improving} our revised  manuscript.
The investigation was carried out with partial support by the ASI grant
no. I/028/12/05.
\\
\\
\\
\newpage
\appendix
\vspace{-0.5cm}
{\fvv As implemented in other contexts of searches for MCAL transients at specific
times provided by external information (e.g., gravitational wave events), we
searched for MCAL triggers in 100 s time windows centered at the $T_0$.

The MCAL burst search logic is based on the criterion that a transient event releases a number of counts above a given threshold over the background. The background count rate depends on the specific timescale and energy range and it is therefore evaluated by adopting different rate meters (RMs): such RMs are characterized by different search integration time (SIT) windows (Hardware trigger logic working on timescales of: 0.293, 1, and 16\,ms; Software trigger logic working on timescales of: 64, 256, 1024, and 8192\,ms), and covering three energy ranges (low energy: $0.3$--$1.4$ MeV; medium energy: 1.4--3 MeV; high energy: 3--100 MeV).

In particular, the hardware 0.293~ms (or ``sub-ms'') search window is a key feature of the \agile MCAL.
Such a feature is crucial for the detection of short-duration events, such as terrestrial gamma-ray flashes (TGFs), lasting few hundreds of microseconds \citep{Marisaldi2010a,2014JGRA..119.1337M}, as well as GRBs with exceptional high fluxes \citep[e.g., GRB 180914B,][]{GCN23226}.

MCAL triggered data acquisitions are issued whenever a given threshold count rate is exceeded. Such thresholds depend on the involved SIT duration. Shortest duration timescales (hardware logic) adopt a static trigger logic, which imposes a fixed threshold count rate $S$. 
{\fvvv For a typical MCAL background rate of 580 Hz, a signal should have at least the following
significances in order to trigger one or more onboard hardware logics \citep[as described in][]{2019ApJ...871...27U}:

\begin{itemize}
\item Sub-ms: at least seven counts on a background of 580 Hz ($\sim$\,0.2 counts/300 $\mu$s), corresponding to $\sim$\,17\,$\sigma$;
\item 1 ms: at least eight counts on a background of 580 Hz ($\sim$\,0.6 counts/1 ms), corresponding  $\sim$\,10\,$\sigma$; and
\item 16 ms: at least 23 counts on a background of 580 Hz ($\sim$\,9 counts/16 ms), corresponding at least at 4\,$\sigma$.
\end{itemize}

The corresponding false alarm rates (FARs) are $\fvvv \sim \,6\times 10^{-4}$\,Hz for the sub-ms,
$\fvvv 3\times 10^{-4}$\,Hz for the 1 ms, and $\fvvv 2\times 10^{-2}$\,Hz for the 16 ms time scales.
These MCAL trigger threshold values were selected mainly for the e.m. counterparts detection
for the LV GW events during the O2 (2017) and O3 (2019--2020) observing runs.
The main goal of this configuration was to increase the total onboard exposure time of
the detector and also to enhance the sensitivity to short-duration impulsive events.
This configuration increases the probability to have weak, short events
falling within the triggered acquisition time windows \citep[as for GW170104; see][]{2017ApJ...850L...27V}.
As a consequence in the case of a candidate coincidence, we are ready to evaluate the c.l. of signals occurring close in time to an external
event according to the post-trial probability of the signal with respect to the external
event time, as in \cite{2016ApJ...826L...6C}. This probability strongly depends  on the
FAR and on the time interval between the signal and the external event, apart from
the signal intrinsic significance above the background.} \\

On the other hand, software logic timescales adopt an adaptive trigger logic, which evaluates the background rate in the given timescale and estimate a related $S=B+N\sigma$ threshold count rate (with $N$ number of standard deviations). Whenever one or more of these thresholds are exceeded, a Burst-START condition is issued and a data acquisition starts. The duration of the data acquisition depends on the Burst-STOP condition, which can be encountered when all RMs go back to a normal background level, or can be forced after a given amount of time (which depends on the timescale). Data are stored in a cyclic buffer, together with pre- and post-burst data streams, regarding time intervals occurring before and after the main burst acquisition. The MCAL trigger logic is flexible and all parameters are fully configurable on ground.\\
}
\newpage


\end{document}